\documentclass[aps,pra, twocolumn,a4paper,showpacs]{revtex4}
\usepackage{amsmath}
\usepackage{bm}
\usepackage{graphicx}

\def\BE {\begin{equation}}
\def\EE {\end{equation}}
\def\BEA {\begin{eqnarray}}
\def\EEA {\end{eqnarray}}
\def\BES {\begin{subequations}}
\def\EES {\end{subequations}}
\def\BA {\begin{array}}
\def\EA {\end{array}}
\def\NN {\nonumber}
\def\pp {^{\prime}}
{

\begin{document}

\title{Field-free two-direction alignment alternation of linear molecules by
elliptic laser pulses}
\author{D.~Daems,$^{1, 2,\star}$ S.~Gu\'erin,$^{1,\star}$ E. Hertz,$^{1}$, H. R.~Jauslin,%
$^{1}$ B. Lavorel $^{1}$ and O. Faucher $^{1,}$}
\email{ddaems@ulb.ac.be, sguerin@u-bourgogne.fr}
\affiliation{$^{1}$Laboratoire de Physique de l'Universit\'e de Bourgogne, UMR CNRS 5027,
BP 47870, 21078 Dijon, France\\
$^{2}$Center for Nonlinear Phenomena and Complex Systems, Universit\'e Libre
de Bruxelles, 1050 Brussels, Belgium}

\begin{abstract}
We show that a linear molecule subjected to a short specific
elliptically polarized laser field yields postpulse revivals
exhibiting alignment alternatively located along the orthogonal
axis and the major axis of the ellipse. The effect is
experimentally demonstrated by measuring the optical Kerr effect
along two different axes. The conditions ensuring an optimal
field-free alternation of high alignments along both directions
are derived.
\end{abstract}

\maketitle

Preparing controlled alignment of molecules is of considerable importance
for a large variety of processes (see \cite{Seideman_review} for a review).
It is well established theoretically and experimentally that the alignment
of a linear molecule along the axis of a linearly polarized field can be of
two types: adiabatic alignment during the interaction with the field, or
transient alignment revivals after a short pulse. The latter is in general
preferred for further manipulations since it offers field-free aligned
molecules. The adiabatic alignment has been extended to three dimensional
alignment of an asymmetric top molecule \cite{Larsen_2000}.

A natural subsequent question was to generate an alignment of a linear
molecule with dynamically varying directions. This question has been studied
using a field of slowly spinning polarization axes which allows one to spin
the axis of alignment and thus the molecule itself \cite{Corkum99}. This
effect, demonstrated experimentally \cite{Corkum00}, has been analyzed using
classical and quantum models \cite{Spanner01}, and in terms of adiabatic
passage through level avoided crossings \cite{Vitanov}. The analysis shows
that the molecule can exhibit a classical rotational motion \textit{while
the field is on}. In this Letter we show a fundamentally different process
in which a linear molecule can dynamically alternate from one direction to
another under field-free conditions. This purely quantum effect is induced
by a suitable short elliptically polarized pulse. The two directions of the
alternation are the major axis and the direction orthogonal to the plane of
the ellipse. The result can be explained using the following qualitative
analysis. It is known that a linear rigid molecule in its ground vibronic
state (of rotational constant $B$) driven by a nonresonant linearly
polarized field (of amplitude $\mathcal{E}$) leads to the Hamiltonian $%
H=H_0+V_{\text{int}}$ with $H_0=BJ^2$, $V_{\text{int}}=-\mathcal{E}%
^2\Delta\alpha\cos^2\theta/4$ (up to a $\theta$-independent constant), the
polarizability anisotropy $\Delta\alpha>0$, and $\theta$ the polar angle
between the field polarization axis and the molecule's axis. This leads to
periodic field-free sequences of revivals that mainly correspond to
alternate alignment along the field axis and planar delocalization \textit{%
orthogonal} to the field axis \cite{Seideman_review}. We emphasize that
unlike the alignment along an axis, the planar delocalization is a specific
quantum effect resulting from the fact that the linear polarization does not
break the planar symmetry orthogonal to the field axis. The effects of alignment and planar delocalization persist after thermal
averaging since, for each molecule, the wave packet produced with different
initial conditions allowed by the thermal Boltzmann distribution keeps the same
periodicity. This has been established theoretically and experimentally \cite%
{Vrakking,Exp_Dijon}. The use of a circular polarization leads to a similar
Hamiltonian: $V_{\text{int}}=\mathcal{E}^2\Delta\alpha\cos^2\theta^{\prime}/8
$ (up to a $\theta$-independent constant) with here $\theta^{\prime}$ the
polar angle between the axis orthogonal to the field polarization and the
molecule axis: the revivals show an alternation between planar
delocalization (in the plane of the field polarization) and alignment along
its \textit{orthogonal} direction \cite{Seideman_review}. Hence, we deduce
that in contrast to the adiabatic case, where the alignment is in the
direction of the minimum of the induced potential ($\theta=0,\pi$ and $%
\theta^{\prime}=\pi/2$ for respectively linear and circular polarization), a
short pulse induces transient alignment (or planar delocalization) in the
directions of the \textit{extrema} of the induced potential. This suggests
that an elliptical polarization can provide an alignment along its major
axis (as a linear polarization would do) and an alignment orthogonal to the
ellipse's plane (as a circular polarization would do). We establish the
validity of this scheme by first calculating explicitly the effective
Hamiltonian. We then identify the optimal parameters of the elliptical
polarization that allow for the alternation of highest alignments between
the two directions. 

We consider a linear (nonpolar) molecule subjected to an elliptically
polarized laser field
\begin{equation}
\overrightarrow{\mathcal{E}}(t)=\mathcal{E}(t)(\overrightarrow{%
e}_{x}a\cos \omega t+\overrightarrow{e}_{y}b\sin \omega t)  \label{pump}
\end{equation}%
of amplitude $\mathcal{E}(t)$, optical frequency $\omega $, and where $a$  represents the half-axis of the ellipse along the $x$-axis
(whereas $b$ corresponds to the $y$-axis) with $a^{2}+b^{2}=1$. When no
excited electronic states and vibrational states are resonantly coupled, the
Hamiltonian is given by \cite{Friedrich95}
\begin{equation}
H=H_{0}-\frac{1}{2}\overrightarrow{\mathcal{E}}(t)\cdot \overrightarrow{%
\overrightarrow{\alpha }}\overrightarrow{\mathcal{E}}(t),
\end{equation}%
with $\overrightarrow{\overrightarrow{\alpha }}$ the dynamical
polarizability tensor which includes the contribution of the excited
electronic states. If we consider frequencies that are low with respect to
the excited electronic states, the dynamical polarizabilities are well
approximated by the static ones. In the limit of high frequency with respect
to the rotation and far from vibrational resonances \cite{Keller}, we obtain
the effective Hamiltonian $H_{\text{eff}}(t)=H_{0}+V_{\text{int}}$ with%
\begin{equation}
V_{\text{int}}=-\frac{\Delta \alpha }{4}\mathcal{E}^{2}(t)\sin
^{2}\theta _{z}\left[ (a^{2}-b^{2})\cos ^{2}\phi _{z}+b^{2}\right]
,  \label{model}
\end{equation}%
with $\phi _{z}$ the azimuthal angle and $\theta _{z}$ the polar
angle, with the choice of the quantum axis along the $z$-axis
orthogonal to the ellipse plane $(x,y)$. Noting that the normalized associated
Legendre functions $\Theta_j^m(\theta)$, i. e. the $\theta$-dependent
part of the spherical harmonics $|j,m\rangle $,   are not necessarily
orthogonal to each other when the sets of indices $(j,m)$ are
different, we obtain
\begin{eqnarray}
\label{I}
&&\left\langle j\pp,m\pp\left| \cos^2 \phi_z \sin^2 \theta_z \right|j,m\right\rangle \NN\\
&&=\delta_{m\pp,m}\left\{A^j_{m}\delta_{j\pp,j}+B_{m}^{j\pp}\delta_{j\pp,j+2}+B_{m}^j\delta_{j\pp,j-2}\right\}\NN\\
&&+ \delta_{m\pp,m+2}\left\{
C_{m}^j\delta_{j\pp,j}+D_{m}^j\delta_{j\pp,j+2}+D^{j\pp}_{-m\pp}\delta_{j\pp,j-2}
\right\} \NN\\
&&+
\delta_{m\pp,m-2}\left\{C_{m\pp}^j\delta_{j\pp,j}+D_{-m}^j\delta_{j\pp,j+2}+D_{m\pp}^{j\pp}
\delta_{j\pp,j-2} \right\} . \quad \end{eqnarray} The coefficients
$A_{m}^j\equiv[1-(c_{m}^j)^2-(c_{m}^{j+1})^2]/2$ and $
B_{m}^j\equiv -c_{m}^{j-1}c_{m}^j/2$  are related to $\theta_z$
only, featuring the standard quantity $c_{m}^j\equiv
[(j-m)(j+m)/(2j-1)(2j+1)]^{1/2}$,   in contrast to the following
coefficients due to both angles: $ C_{m}^j\equiv
-[(j-m)(j-m-1)(j+m+2)(j+m+1)]^{1/2}/2(2j-1)(2j+3)$ and $
D_{m}^j\equiv[(j+m+4)!/(2j+1)(2j+5)(j+m)!]^{1/2}/4(2j+3)$. We  use
a laser pulse of short duration  that can be treated in the sudden
approximation, where  the intensity of the field is characterized
by the dimensionless parameter \cite{Henriksen,David} $
\xi=\frac{\Delta\alpha}{4\hbar}\int dt {\cal E}^2(t) $.
\begin{figure}[h!]
{\includegraphics[scale=0.73]{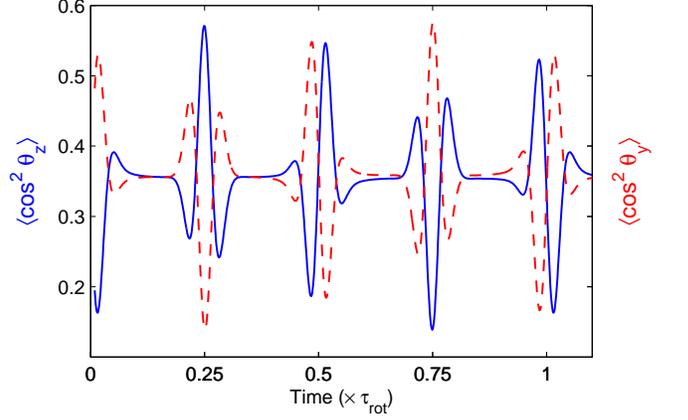}}
\caption{(Color online) Expectation values of the observables $\cos^2\protect%
\theta_z$ (full line) and $\cos^2\protect\theta_y$ (dashed line)
for $a^2=1/3$, $\protect\xi=11.1$ and $%
\tilde T=20$, as a function of normalized time.}
\label{FigT}
\end{figure}

\begin{widetext}
\begin{figure}[t] {\includegraphics[scale=0.73]{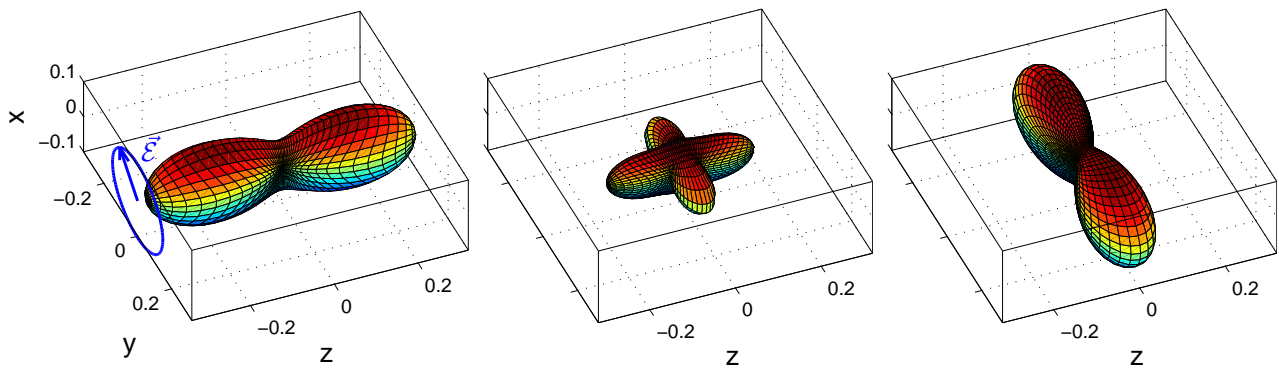}}
\caption{(Color online) Representation of the molecular state in
spherical coordinates
$\left\{r\equiv|\psi(\theta_z,\phi_z;t)|^2,\theta_z,\phi_z\right\}$
for $a^2=1/3$, $\xi=11.1$, $\tilde T=20$, at times $t=\tau_{\text
{rot}}/4$ (left panel), $t\approx 3\tau_{\text {rot}}/8$
(middle panel) and $t=3\tau_{\text {rot}}/4$ (right panel). The polarization ellipse 
 of the field $\vec{\mathcal{E}}$ is sketched in the
$(x,y)$ plane.} \label{Fig3D}
\end{figure}
\end{widetext}

To analyze the alignments along the two axes, in addition to   $\langle \cos ^{2}\theta _{z}\rangle (t)\equiv \langle \psi
(\theta _{z},\phi _{z};t)|\cos ^{2}\theta _{z}|\psi (\theta _{z},\phi
_{z};t)\rangle $ one can consider the
observable $\langle \cos ^{2}\phi _{z}\rangle (t)\equiv \langle
\psi (\theta _{z},\phi _{z};t)|\cos ^{2}\phi _{z}|\psi (\theta _{z},\phi
_{z};t)\rangle $, where $|\psi (\theta _{z},\phi _{z};t)\rangle $ is the
state of the molecule given by the Schr\"{o}dinger equation. However, it is more appropriate  to introduce  the observables $\langle \cos
^{2}\theta _{x}\rangle (t)\equiv \langle \cos ^{2}\phi _{z}\sin ^{2}\theta
_{z}\rangle (t)$ and $\langle \cos ^{2}\theta _{y}\rangle (t)\equiv \langle
\sin ^{2}\phi _{z}\sin ^{2}\theta _{z}\rangle (t)$, $\theta _{x}$ ($\theta
_{y}$) corresponding to the polar angle with respect to the $x$-axis ($y$%
-axis).  This is motivated by the fact that these observables are closely
related to the experimental measurements and the fact  that they will allow us to identify the ellipticity leading to quantitatively equivalent alignments along both the major axis ($x$ or $y$) of the polarization ellipse  and  the $z$-axis orthogonal to the ellipse. Because of the relation $\sum_{i=x,y,z}\langle \cos ^{2}\theta _{i}\rangle
(t)=1$, the alignment alternation can be measured with any pair of observables among $\{\langle \cos
^{2}\theta _{i}\rangle (t),\ i=x,y,z\}$.
 Figure \ref{FigT}
displays the temporal behaviour of these thermally averaged
quantities for an elliptically polarized field interacting with a
molecule at the dimensionless temperature $\tilde{T}:=kT/B=20$ (an
initial statistical ensemble of even values of $j$ is considered).
This amounts to having $T=11$ K for a CO$_{2}$ molecule. Here
$a^2=1/3$ and $\xi =11.1$ (corresponding approximately to a pulse
of peak
intensity $I=25\times 10^{12}$ W/cm$^{2}$ and of duration $\tau _{\text{FWHM}%
}=100$~fs for CO$_{2}$). During each rotational period $\tau
_{\text{rot}}$ we can identify four revivals for both $\langle
\cos ^{2}\theta _{y}\rangle (t)$ and $\langle \cos ^{2}\theta
_{z}\rangle (t)$. The revivals occur around the times $t_{n}=n\tau
_{\text{rot}}/4$ for both expectation values, as is the case for a
linear polarization. The localization properties of the rotational
wave packet are however fundamentally different. Near the highest peaks of
$\langle \cos ^{2}\theta _{z}\rangle (t)$ (at the times $t_{1}$, slightly after $t_{2}$ and also slightly before $t_{4}$), the
molecule is predominantly aligned along the $z$-direction (small $\theta _{z}$), i. e. orthogonally to the
polarization ellipse. This
state is represented in spherical coordinates at time $t=t_{1}\equiv \tau _{%
\text{rot}}/4$ in Fig.~\ref{Fig3D} (left panel). At the highest peaks of  $\langle \cos ^{2}\theta _{y}\rangle (t)$ (slightly before $t_{2}$, at
the time $t_{3}$ and also slightly after
$t_{4}$), coinciding with the  minima of
$\langle \cos ^{2}\theta _{z}\rangle (t)$, the molecule is aligned along the major axis (small $\theta _{y}$ and $\theta
_{z}$ close to $\pi /2$).
A representation of the molecular state is displayed at time $%
t=t_{3}\equiv 3\tau _{\text{rot}}/4$ in Fig.~\ref{Fig3D} (right panel).

As will be discussed below, 
 the alignments
revivals are quantitatively similar in both the $y$ and $z$-directions for the particular value $a^2=1/3$.
The quantities 
 $\langle \cos ^{2}\theta _{z}\rangle (t)$ and $\langle
\cos ^{2}\theta _{y}\rangle (t)$ displayed in Fig. \ref{FigT}  are   symmetric with respect to the approximate value 0.36, implying that $\langle \cos ^{2}\theta _{x}\rangle (t)$ is close to 1/3 for all
times.
 The angular distributions shown in Fig. \ref{Fig3D} for $t=\tau_{\text {rot}}/4$ and $t=3\tau_{\text {rot}}/4$  are superposable upon rotation.

Between the revivals, when the averaged observables are approximately flat
as a function of time, locally near the times $t=(2p+1)\tau_{\text{rot}}/8$
(with integer $p$), it is remarkable that the state of the molecule is
approximately an equal weight superposition of the two aligned states, as
illustrated in Fig.~\ref{Fig3D} (middle panel). This can be interpreted as
an extension of a recent proposal made in the linearly polarized case \cite%
{Spanner}, where fractional revivals at odd multiples of
$\tau_{\text{rot}}/8 $ are shown to combine aligned (along the
linear polarization axis) and antialigned (delocalized in the
plane orthogonal to the axis) components with equal weights. In
the current elliptic case, both components correspond to aligned
states.

So far we have considered $a^2=1/3$. We now turn to the question
of determining the value of this parameter giving an optimal two-direction
alignment alternation, in the sense that both alignments correspond to
similar (i. e. superposable upon rotation) delocalized angular distribution. Choosing
$a= a_{\text{lin}}\equiv0$ ($a=1$) gives a linear polarization along the $y$%
-axis ($x$-axis).
\begin{figure}[h!]
{\includegraphics[scale=0.73]{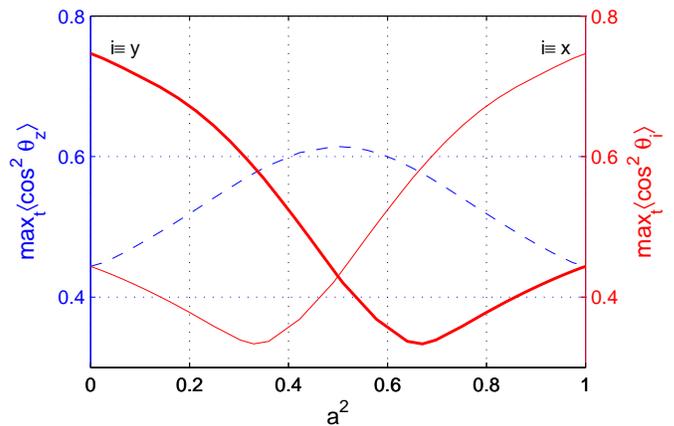}} \caption{(Color online)
Maximum over time
of $\langle\cos^2\protect\theta_z\rangle$ (dashed line), $\langle\cos^2%
\protect\theta_y\rangle$ (full thick line), and $\langle\cos^2%
\protect\theta_x\rangle$ (full thin line) with $\protect\xi=11.1$ and
$\tilde T=20$, as a function of $a^2$.} \label{figa}
\end{figure}
The circular polarization is obtained with $a^2=b^2=
a_{\text{circ}}^2\equiv1/2$. The optimal value is obtained when
the maxima (over time) of both expectation values $\langle\cos^2\protect\theta_z\rangle$
and $\langle\cos^2\protect\theta_y\rangle$ for $a^2<1/2$ 
($\langle\cos^2\protect\theta_x\rangle$ for $a^2>1/2$) are equal.
In Fig.~\ref{figa} we plot these maxima
as a function of $a^2$.
The intersection points are near $a^2 =
1/3$ (corresponding to the ellipticity chosen for Figs \ref{FigT}
and \ref{Fig3D}) and $a^2 = 2/3$. 
This can be understood by rewriting
the interaction term (\ref{model}) in terms of the observable $\cos^2 \theta_y$ rather than   $\cos^2 \phi_z$:
\begin{equation}
V_{\text{int}} =-\frac{\Delta \alpha }{4}\mathcal{E}^{2}(t)\left[
(1-2a^{2})\cos ^{2}\theta _{y}+a^{2}\sin ^{2}\theta
_{z}\right].
\label{model2}
\end{equation} 
When the ellipticity is chosen such that
$1-2a^2=a^2$, i.e. $a^2=1/3$,  the
directions $\theta_y$ and $\theta_z$ play a symmetric role.
Notice from (\ref{model2}) that the minima in one direction correspond to the maxima in the other direction.
In the sudden regime, both types of extrema are visited equivalenty in  contrast to the adiabatic case.
The ellipticity $a^2=1/3$ can be interpreted as the best compromise
between linear and circular polarizations with the remarkable
feature that the angular distribution associated with the highest
revivals are similar in both directions (see Fig.
\ref{Fig3D}). It is worth noting that this value is not the arithmetic average of $a_{\text{circ}}^2$ and  $a_{\text{lin}}^2$. The case $a^2=2/3$ involves the direction $\theta_x$ instead of $\theta_y$. In Fig.~\ref{figa} we also recognize the case of a
circular polarization for $a^2=1/2$ where the
alignment along $z$ is large whereas the maxima of $%
\langle\cos^2\theta_x\rangle$ and $\langle\cos^2\theta_y\rangle$
are equal, reflecting the fact that the rotational wavepacket is
delocalized for all times in the $x,y$-plane.

\begin{figure}[h!]
{\includegraphics[scale=0.73]{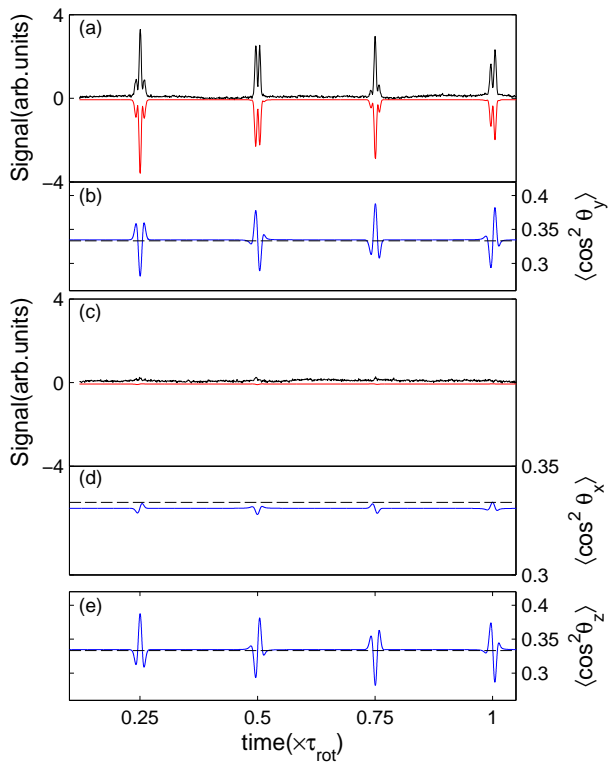}} \caption{(Color online)
Field-free alignment of an ensemble of CO$_2$ molecules
($T=296$~K, $P=4\times10^4$~Pa) produced by an elliptically
polarized pump field [$a^2\approx1/3$, see
Eq.~(\protect\ref{pump})] of peak
intensity 25 TW/cm$^2$ and pulse duration $\protect\tau_{\text{FWHM}}=100$%
~fs. Cross defocusing signal recorded as function of delay with the probe
field linearly polarized along (a) the $y$-axis (major field axis) and (c)
the $x$-axis (minor field axis). In (a) and (c), the theoretical results  (thick lines) are
shown as mirror images of the experimental ones. Corresponding observables (b) $\langle\cos^2\protect%
\theta_y\rangle$ and (d) $\langle\cos^2\protect\theta_x\rangle$. (e) $%
\langle\cos^2\protect\theta_z\rangle$ along the propagation
$z$-axis deduced from (b) and (d). The isotropic values 1/3 are
indicated with dashed lines.} \label{exp}
\end{figure}

We have demonstrated the effect experimentally in CO$_2$ molecules
at room temperature by measuring the optical Kerr effect along the
two orthogonal directions, $x$ and $y$ respectively (both
orthogonal to the $z$-direction of propagation of the beam). The
measurements have been performed with a Ti:sapphire chirped-pulse
amplifier producing 100-fs pulses at 1~kHz. Recently it has been
shown \cite{Renard05} that measuring the defocusing of a
time-delayed weak probe pulse produced by a spatial distribution
of aligned linear molecules yields a signal proportional to
$(\langle\cos^2\theta\rangle(t)-1/3)^2$, with $\theta$ the angle
between the molecular axis and the direction of the probe field.
Choosing the polarization of the probe either in the $y$ or $x$-direction,
we can thus obtain $[\langle\cos^2\theta_i\rangle(t)-1/3]^2$ (shown in Fig. %
\ref{exp}a,c), where $\theta_i$ is the angle between the molecular
axis
and the $i$-axis ($i=x,y$). From these we can deduce $\langle\cos^2\theta_y%
\rangle$ and $\langle\cos^2\theta_x\rangle$ which characterize the alignment
along the major and minor axes of the ellipse, respectively (Fig. \ref{exp}b, d). The shape and amplitude of the recorded signal are in good
agreement with the theoretical predictions. The quasi-isotropic
feature of the alignment along the $x$-axis mentioned above for
this specific ellipticity is confirmed by this experiment, where a
signal close to zero is observed in Fig. \ref{exp}c.
The alignment along the  $z$-axis , characterized by $%
\langle\cos^2\theta_z\rangle(t)$ (Fig. \ref{exp}e), can be deduced
from the other two observables through the relation
$\sum_{i=x,y,z}\langle\cos^2\theta_i\rangle(t)=1$. The results of
Fig. \ref{exp}b and e show clearly the
experimental alternation of the alignment predicted theoretically and represented in Fig.~%
\ref{FigT}. It should be noted that the experimental signal
related to the measurement of
$(\langle\cos^2\theta_x\rangle(t)-1/3)^2$ is found to be minimum for
the ellipticity $a^2=1/3$, as predicted by the model. As discussed above, the fact that $\langle\cos^2\theta_x\rangle(t)\approx1/3$ (in a strong
field) is a clearcut signature of optimal alignments in the two
other directions. An exhaustive experimental investigation with
other ellipticities has been performed and shows strong
modifications of both shape and amplitude of the recorded signals, in agreement with numerical simulations.

In conclusion, we have shown, both theoretically and experimentally, that a linear molecule subjected to a
short specific elliptically polarized laser field can be aligned,
alternatively, at specific times along the orthogonal axis and the
major axis of the ellipse. Contrary to the adiabatic case where
only the minima of the induced potential are populated, for short
pulses all the extrema of the potential are dynamically visited
and appear as revivals. The control of this field-free
two-direction alignment alternation is a challenging perspective
that could find applications in nano-technology, for instance to
generate a 3D molecular switch \cite{Seideman_review}. In the context of quantum
information, the advantage of an elliptic polarization over a
linearly polarized field employed in a recent proposal \cite{Lee}
is to have a superposition of alignments along two axes, instead
of a superposition of an alignment and a planar delocalization.

This research was supported by the \textit{Conseil R\'{e}gional de Bourgogne}%
, the \textit{Action Concert\'{e}e Incitative Photonique} from the French
Ministry of Research, and a Marie Curie \textit{European Reintegration Grant}
within the 6th European Community RTD Framework Programme.

\end{document}